\documentstyle[amssymb,prl,aps,twocolumn,epsfig]{revtex}

\begin{document}

\twocolumn[\hsize\textwidth\columnwidth\hsize\csname @twocolumnfalse\endcsname

\title{Compressibility and Electronic Structure of MgB$_2$ up to 8 GPa}
\author{T. Vogt and G. Schneider}
\address{Physics Department,Brookhaven National Laboratory,Upton, NY 11973-5000,USA}
\author{J.A. Hriljac}
\address{School of Chemistry, The University of Birmingham, Edgbaston, Birmingham B152TT, United Kingdom}
\author{G. Yang and J.S. Abell}
\address{School of Metallurgy and Materials, The University of Birmingham, Edgbaston, Birmingham B152TT, United Kingdom}
\date{\today}
\maketitle

\begin{abstract}

The lattice parameters of MgB$_2$ up to pressures of 8 GPa were determined using
high-resolution x-ray powder diffraction in a diamond anvil cell.
The bulk modulus, $B_0$, was determined to be $ 151\pm 5$ GPa. 
Both experimental and first-principles calculations indicate nearly isotropic
mechanical behavior under pressure. This small anisotropy is in contrast to
the 2 dimensional nature of the boron $p$ states.
The pressure dependence of the density of states at the Fermi level 
and a reasonable value for the average phonon frequency 
$\langle\omega\rangle$ account within the context of BCS theory for the 
reduction of $T_c$ under pressure.

PACS numbers: 
\end{abstract}

\pacs{}
\vskip1pc]

\narrowtext

The recent discovery of superconductivity in MgB$_{2}$ close to 40K \cite
{aki} has pushed $T_{c}$ beyond what was thought to be possible in
intermetallic compounds within the context of BCS theory \cite
{McMillan,Cohen}. However, the boron isotope effect was measured to be
consistent with MgB$_{2}$ being a conventional phonon-mediated BCS
superconductor \cite{Budko}.The temperature-dependent electrical resistivity
in the normal state was observed to obey a $T^{3}$ power law by Finnermore,
et al. \cite{Finnermore}, whereas Jung et al. \cite{Jung} found a $T^{2}$
behavior to be more appropriate. Within the context of hole superconductivity 
\cite{Hersch} it was proposed that a decreasing B-B intraplane distance
should increase $T_{c}$. However, ac magnetic susceptibility measurements
revealed that $T_{c}$ actually decreases \cite{lorenz01}. The response of the
crystal and electronic structures of MgB$_{2}$ to pressure is not only
important to distinguish a conventional from a non-BCS mechanism, but it
might also give valuable clues for guiding chemical substitutions. Slusky, et
al. \cite{Slusky} have established the loss of superconductivity in Mg$_{1-x}$%
Al$_{x}$B$_{2}$ and conjecture that MgB$_{2}$ is on the edge of a structural
instability at slightly higher electron concentrations.These authors
furthermore indicate the presence of a two phase region between 
$0.1 \lessapprox x\lessapprox 0.25$. To investigate the intrinsic electronic
properties of MgB$_{2}$ and to avoid complications from possible extrinsic
effects such as compositional fluctuations and phase separations 
in substituted systems, we have
determined the bulk compressibility and compared it to the one derived from
first-principles calculations based on Density Functional Theory (DFT).

The sample, obtained from Alfa, was of excellent quality without any
impurities visible in a high-resolution x-ray powder diffraction pattern
taken at room temperature. All reflections could be indexed to the AlB$_{2}$%
-type structure in spacegroup P6/mmm with a=3.08589(1) \AA\ and c=3.52121(3)
\AA\ at room temperature and ambient pressure. The magnetic susceptibility
showed a sharp transition at 38.2K (Figure 1). The compressibility
up to 8 GPa was measured using a Merrill-Bassett diamond anvil cell. The
lattice parameters (Table 1) were determined using the high-resolution
powder diffractometer X7A at the National Synchrotron Light Source at
Brookhaven National Laboratory. An asymmetrically cut triangular shaped Si
(220) monochromator was cylindrically bent to focus a photon beam with a
wavelength of 0.68613(2) \AA\ down to 200 $\mu $m. A steel gasket was
pre-indented and a 200 $\mu $m hole drilled in the center. The sample and
several small ruby chips were loaded along with the pressure transmission
fluid of methanol:ethanol:water (ratios 16:3:1). Pressure was determined by
measuring the shift of the R1 emission line of ruby using an argon ion laser 
\cite{Piermarini}. Diffraction patterns up to $2\theta =35^{\circ}$ were
recorded using a position-sensitive detector \cite{Smith}. Further
experimental details are described in \cite{Jephcoat}. There were no
indications of any splitting or significant broadening of the Bragg
reflections indicative of a possible phase transition. The width of the R1
emission line of ruby at higher pressures also indicated that the sample was
well within a hydrostatic pressure regime. The unit cell volumes, $V$,
normalized to the one at ambient pressure, $V_{o}$, at various pressures $P$
were fitted to a first-order Murnaghan equation of state $%
V=V_{0}(1+B_{0}^{\prime }P/B_{0})^{-1/B_{o}{}^{\prime }}$where $B_{0}$ is
the bulk modulus at ambient conditions and $B_{0}\prime =4$. We obtained a
value of $151\pm 5$ GPa. For comparison, $B_{0}$ of YNi$_{2}$B$_{2}$C, a
related borocarbide, was measured to be 200 GPa \cite{Meenakshi}.

We calculated the electronic structure and mechanical properties within
DFT\cite{hoko6465} using the full-potential linearized
augmented plane wave method FLAPW\cite{flapw}. For the exchange-correlation
potential we used the Generalized Gradient Approximation of Perdew, et al. 
\cite{rpbe98}. A relatively large basis set was required to accurately
calculate small changes in the electronic structure and lattice parameters
with applied pressure\cite{lapwparam}. Calculated equilibrium parameters are 
$a=3.089$\AA, $c=3.548$\AA\ and $c/a=1.149$. They differ from the experimental
values by 0.1\%, 0.8\% and 0.7\% respectively. The calculated bulk modulus
of $B_{0}=139\pm 10$ GPa is in good agreement with the measured value of $%
151\pm 5$ GPa. Both calculated and experimental lattice parameters as a 
function of volume are shown in Fig. 2. The agreement between theory and 
experiment is excellent. Over the range of pressure considered, the $c/a$ ratio
is essentially constant, indicative of the 3 dimensional character of MgB$_2$.
The small observed anisotropy
of the compressibility as given by the observed reduction in the $c/a$ ratio
with decreasing volume is reproduced by our calculations, although the rate 
of change is overestimated.

The superconducting properties of MgB$_{2}$ as a function of pressure have
been investigated by Lorenz, et al. \cite{lorenz01}, who found that the
transition temperature $T_{c}$ decreases linearly and reversibly at a rate
of $-1.6$ K/GPa. Within BCS theory the pressure dependence of $T_{c}$ can be
calculated from the the McMillan formula 
\[
T_{c}=\langle \omega \rangle \exp \left( \frac{-(1+\lambda )}{\lambda -\mu
^{\ast }-\lambda \mu ^{\ast }}\right) ,
\]
where $\langle \omega \rangle $ is an average phonon frequency, $\mu ^{\ast }
$ denotes the Coulomb pseudopotential, $\lambda \propto N(0)/\langle \omega
^{2}\rangle $ is the 
electron-phonon interaction parameter and $N(0)$ denotes the density of
states at the Fermi energy\cite{McMillan}. 

The calculated change in the density of states of MgB$_{2}$ for a pressure
of $\approx 8$~GPa is shown in Figure 3. We observe the expected
increase in bandwidth, but the overall changes are very small and in
particular $N(0)$ decreases by
only 3\% from 0.70 states/eV to 0.68 states/eV. Assuming that $T_{c}$
continues to decrease linearly at higher pressures we can estimate that at
8 GPa, $T_{c}\approx 26K$ which is 30\% lower than $T_{c}$ at ambient
pressure. Comparing this large decrease in $T_{c}$ to the minute change in $%
N(0)$, it becomes apparent that the pressure dependence of $N(0)$ alone
cannot account for the large rate with which $T_{c}$ decreases with applied
pressure. Taking our value for $d \ln N(0)/dP \approx -0.004/$GPa and
assuming $\mu^{*} = 0.1$ and $\lambda=1.0$, which is the average of
published calculations and estimates ranging from 0.7 to 1.4 
\cite{kortus01,anpi01}, we find that in order to reproduce the experimentally
observed pressure dependence of $T_c$, $d \ln \langle\omega^2\rangle^{1/2} / dP
\approx +0.014/$ GPa is needed. This value appears reasonable.

The volume dependence of $T_{c}$ within BCS theory has been characterized
using 
\[
\frac{d\ln (T_{c}/\Theta_D)}{d \ln V} =  \ln \frac{\Theta_D}{T_{c}} \frac{%
d\ln \lambda }{d \ln V} \equiv  \ln \frac{\Theta_D}{T_{c}} \varphi 
\]
where $\varphi$ is a material dependent parameter describing the volume 
dependence of the electron-phonon coupling parameter and $\Theta_D$ is the 
Debye temperature.
For superconducting
$sp$ metals one generally finds $\varphi \approx 2.5$\cite{levy64}.
Using for the Debye temperature the recently determined value of $\theta_D
= 800$~K \cite{kremer01}, we find $\varphi = 2.4$ for MgB$_2$, a value not
inconsistent with BCS theory.

The importance of the mostly 2-dimensional B $p_{xy}$ states for the
superconductivity of MgB$_2$ has been pointed out by several authors 
\cite{kortus01,anpi01}. The change with pressure in the partial density of
states (PDOS) of B $p$ states separated into $p_{xy}$ and $p_z$ character is
shown in Fig. 4. The B $p_z$ PDOS remain essentially unchanged, but the B 
$p_{xy}$ PDOS shows an overall shift to lower energies, leading to a loss of
states in the energy range from just below the Fermi energy to $\approx
-2$ eV. In general one expects the PDOS inside a fixed sphere to increase
with pressure.The qualitatively different behavior of the B $p_{xy}$ PDOS in
MgB$_2$ can be attributed to the 2-dimensional character of these B $p_{xy}$
states. However, the B $p_{xy}$ PDOS at the Fermi energy does not change
significantly.

In conclusion, we have experimentally and theoretically determined 
the bulk modulus of 
MgB$_2$ and lattice parameters up to pressures of 8 GPa. Despite an overall 
isotropic behavior of MgB$_2$, there are small anisotropies in the mechanical
properties.
The electronic structure under pressure reveals the
2-dimensional character of the B $p_{xy}$ states whose occupancy is altered by 
pressure in contrast to the B $p_{z}$ states.

Furthermore, the pressure dependence of 
the density of states at the Fermi level 
and a reasonable value for the average phonon frequency 
$\langle\omega\rangle$ account for the 
reduction of $T_c$ under pressure within the context of BCS theory.

We find that the material dependent constant $\varphi$ describing
the volume dependence of the electron-phonon coupling parameter agrees well with
BCS theory for an $sp$ metal.

We gratefully acknowledge discussions with Mike Weinert and Myron Strongin. We
also would like to thank J. Hu from the Geophysical Laboratory, Carnegie
Institution of Washington for being able to use the laser system for
pressure determination. The work was supported by the Division of Materials
Science, US Department of Energy, under Contract NO. DE\_AC02-98CH10886 and
a grant of computer time at the National Energy Research Scientific
Computing Center (NERSC).



\begin{references}
\bibitem{lapwparam}  The parameters used in all calculations are $R_{MT} =
1.6 (2.7)$ a.u. for B (Mg) atoms. Potential and charge density were expanded
up to $l_{max}=8$ and $G_{max}=18\,a.u.^{-1}$. The wave-function cutoff was
chosen to be $R_{MT}k_{max}=8.8$ leading to $\approx 180$ basis functions
per atom.
\end{references}


\newpage

\begin{table}[tbp]
\caption{Measured lattice parameters for MgB$_2$ as a function of applied
pressure. The error in pressure is estimated to ~0.2 GPa.}
\label{table2}
\begin{tabular}{cccccccc}
pressure (GPa) & volume (\AA$^3$) & $a$ (\AA) & $c$ (\AA) & $c/a$ &  &  & 
\\ 
\tableline 0.0 & 29.0391 & 3.08589(3) & 3.52121(9) & 1.1411 &  &  &  \\ 
1.17 & 28.8494 & 3.08017(28) & 3.51121(29) & 1.1399 &  &  &  \\ 
2.14 & 28.5835 & 3.07149(21) & 3.49854(52) & 1.1390 &  &  &  \\ 
3.05 & 28.5017 & 3.06709(49) & 3.48857(58) & 1.1374 &  &  &  \\ 
4.07 & 28.2994 & 3.06349(15) & 3.48188(25) & 1.1366 &  &  &  \\ 
5.09 & 28.0519 & 3.05449(19) & 3.47180(24) & 1.1366 &  &  &  \\ 
6.53 & 27.8583 & 3.04972(31) & 3.45863(24) & 1.1341 &  &  &  \\ 
8.02 & 27.8228 & 3.04843(14) & 3.45715(28) & 1.1341 &  &  & 
\end{tabular}
\end{table}

\begin{figure}[tbp]
\epsfig{width=0.85 \linewidth,figure=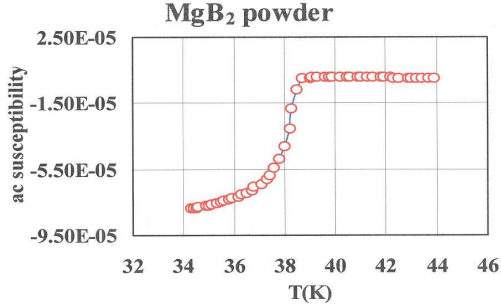}
\caption{ac susceptibility of MgB$_2$ as a function of temperature.}
\end{figure}

\begin{figure}[tbp]
\epsfig{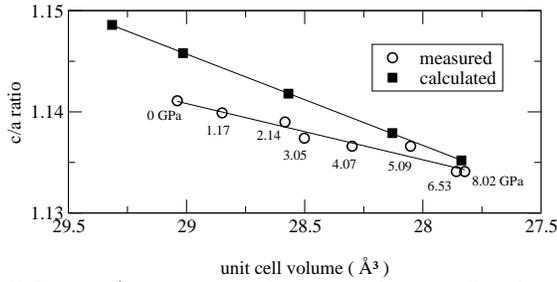}
\caption{c/a ratio as a function of unit-cell volume from both experiment
and calculation.}
\end{figure}

\begin{figure}[tbp]
\epsfig{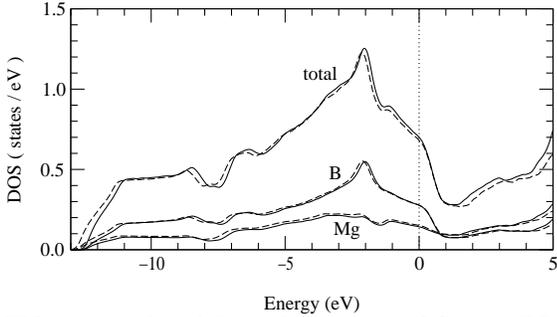}
\caption{Total and Partial Densities of States (PDOS) of MgB$_2$ for two
sets of lattice parameters: ( $a=3.09$~\AA, $c=3.55$~\AA ) (full lines) and
(  $a=3.05$~\AA, $c=3.46$~\AA ) (dashed lines). The difference in lattice
parameters corresponds to a pressure of $\approx 8$~GPa. The energy zero is
the Fermi energy. The PDOS is calculated inside the Muffin tin spheres ( $%
R_{MT}=1.6$~a.u. and 2.7~a.u. for B and Mg ). }
\end{figure}

\begin{figure}[tbp]
\epsfig{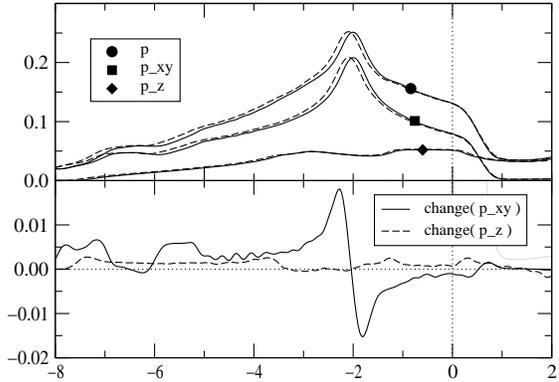}
\caption{Single Boron p Partial Density of States (PDOS) (top panel) for two
sets of lattice parameters (see Figure 3), corresponding to 0~Gpa
(full lines) and $\approx 8$~GPa (dashed lines). The change in the B $p_{xy}$
PDOS (full line) and $p_{z}$ PDOS (dashed line) is shown in the bottom panel. 
The energy zero is the Fermi energy.}
\end{figure}

\end{document}